\documentclass[aps,prd,twocolumn,showpacs,amsmath,amssymb]{revtex4-1}
\usepackage{amsmath} \usepackage{graphicx} \usepackage{subfigure}
\usepackage{epstopdf} \usepackage{color} \usepackage{multirow}
\usepackage{setspace} \usepackage{overpic} \usepackage{amssymb}
\usepackage{float}
\usepackage{amsmath}
\usepackage[bookmarksnumbered, pdfstartview=FitH,colorlinks,urlcolor=blue, citecolor=blue,linkcolor=blue] {hyperref}
\usepackage{lineno}
\usepackage{bm}
\usepackage{rotating}
\usepackage{xcolor}
\usepackage{makecell}
\usepackage{mathtext}
\usepackage{mathrsfs}
\usepackage[utf8]{inputenc}

\newcommand{\BESIIIorcid}[1]{\href{https://orcid.org/#1}{\hspace*{0.1em}\raisebox{-0.45ex}{\includegraphics[width=1em]{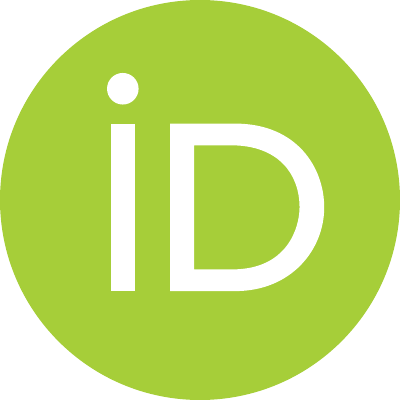}}}}

\let\oldequation\equation
\let\oldendequation\endequation
\renewenvironment{equation}
 {\linenomathNonumbers\oldequation}
 {\oldendequation\endlinenomath}

\definecolor{boslv}{rgb}{0.0, 0.65, 0.58}
\definecolor{Munsell}{HTML}{00A877}

\newcommand{\Br}{\mathcal{B}}
\newcommand{\mL}{\mathcal{L}}
\newcommand{\too}{\rightarrow}

\newcommand{\LLB}{\Lambda\bar{\Lambda}}

\newcommand{\pip}{\pi^+}
\newcommand{\pim}{\pi^-}
\newcommand{\pizero}{\pi^0}

\newcommand{\Xizero}{\Xi^{0}}
\newcommand{\Xip}{\bar{\Xi}^{+}}
\newcommand{\Xim}{\Xi^{-}}
\newcommand{\Xizerob}{\bar{\Xi}^{0}}

\newcommand{\EE}{e^+e^-}

\newcommand{\bfg}{\begin{figure}}
\newcommand{\efg}{\end{figure}}
\newcommand{\bitm}{\begin{itemize}}
\newcommand{\eitm}{\end{itemize}}
\newcommand{\bnum}{\begin{enumerate}}
\newcommand{\enum}{\end{enumerate}}
\newcommand{\btbl}{\begin{table}}
\newcommand{\etbl}{\end{table}}
\newcommand{\btbu}{\begin{tabular}}
\newcommand{\etbu}{\end{tabular}}
\newcommand{\bcl}{\begin{center}}
\newcommand{\ecl}{\end{center}}

\newcommand{\beq}{\begin{equation}}
\newcommand{\eeq}{\end{equation}}
\newcommand{\beqr}{\begin{eqnarray}}
\newcommand{\eeqr}{\end{eqnarray}}

\newcommand{\blue}{\color{blue}}

\begin{document}

\title{\boldmath Experimental study of the reaction $\Xi^{0}n\too\Lambda\Lambda X$ using $\Xi^{0}$-nucleus scattering}

\author{
\begin{small}
\begin{center}
M.~Ablikim$^{1}$\BESIIIorcid{0000-0002-3935-619X},
M.~N.~Achasov$^{4,c}$\BESIIIorcid{0000-0002-9400-8622},
P.~Adlarson$^{77}$\BESIIIorcid{0000-0001-6280-3851},
X.~C.~Ai$^{82}$\BESIIIorcid{0000-0003-3856-2415},
R.~Aliberti$^{36}$\BESIIIorcid{0000-0003-3500-4012},
A.~Amoroso$^{76A,76C}$\BESIIIorcid{0000-0002-3095-8610},
Q.~An$^{59,73,a}$,
Y.~Bai$^{58}$\BESIIIorcid{0000-0001-6593-5665},
O.~Bakina$^{37}$\BESIIIorcid{0009-0005-0719-7461},
Y.~Ban$^{47,h}$\BESIIIorcid{0000-0002-1912-0374}
% ...
\\
\vspace{0.2cm}
(BESIII Collaboration)\\
\vspace{0.2cm} {\it
$^{1}$ Institute of High Energy Physics, Beijing 100049, People's Republic of China\\
% ...
$^{4}$ Budker Institute of Nuclear Physics SB RAS (BINP), Novosibirsk 630090, Russia\\
% ...
$^{36}$ Johannes Gutenberg University of Mainz, Johann-Joachim-Becher-Weg 45, D-55099 Mainz, Germany\\
$^{37}$ Joint Institute for Nuclear Research, 141980 Dubna, Moscow region, Russia\\
% ...
$^{47}$ Peking University, Beijing 100871, People's Republic of China\\
% ...
$^{58}$ Southeast University, Nanjing 211100, People's Republic of China\\
$^{59}$ State Key Laboratory of Particle Detection and Electronics, Beijing 100049, Hefei 230026, People's Republic of China\\
% ...
$^{73}$ University of Science and Technology of China, Hefei 230026, People's Republic of China\\
% ...
$^{76}$ University of Turin and INFN, (A)University of Turin, I-10125, Turin, Italy; (B)University of Eastern Piedmont, I-15121, Alessandria, Italy; (C)INFN, I-10125, Turin, Italy\\
$^{77}$ Uppsala University, Box 516, SE-75120 Uppsala, Sweden\\
% ...
$^{82}$ Zhengzhou University, Zhengzhou 450001, People's Republic of China\\
% ...
$^{a}$ Deceased\\
% ...
$^{c}$ Also at the Novosibirsk State University, Novosibirsk, 630090, Russia\\
% ...
$^{h}$ Also at State Key Laboratory of Nuclear Physics and Technology, Peking University, Beijing 100871, People's Republic of China\\
% ...
}
\end{center}
\vspace{0.4cm}
\end{small}
}


\begin{abstract}
 Using $(10087\pm44)\times10^{6}$$J/\psi$ events collected with the BESIII detector operating at the BEPCII storage ring in $2009$, $2012$, $2018$, and $2019$, we perform a search for the reaction $\Xi^0n\too\Lambda\Lambda X$, where $X$ denotes any additional final particles. Given the highly suppressed phase space for producing extra pions, the $X$ consists of either nothing or a photon, corresponding to the processes $\Xi^0 n \to \Lambda\Lambda$ and $\Xi^0 n \to \Lambda\Sigma^0 \to \Lambda\Lambda\gamma$. The $\Xi^0$ comes from the decay of $J/\psi\too\Xi^0\bar{\Xi}^0$, while the neutron originates from material of the beam pipe. A signal is observed for the first time with a statistical significance of 6.4$\sigma$. The cross section for the reaction $\Xi^0+{^9\rm{Be}}\too\Lambda+\Lambda+X$ is measured to be $(43.6\pm10.5_{\text{stat}}\pm11.1_{\text{syst}})$~mb at $P_{\Xi^0}\approx0.818$ GeV/$c$, where the first uncertainty is statistical and the second systematic. No significant $H$-dibaryon signal is observed in the $\Lambda\Lambda$ final state.
\end{abstract}

\maketitle

Scattering experiments with high-energy particle beams incident on target materials have been of great significance for studying the inner structure of matter and the fundamental interactions~\cite{cite1,cite2,cite3}. However, due to their significantly shorter lifetimes and higher masses, hyperon beams, such as those of $\Lambda$, $\Sigma$, or $\Xi$, are more difficult to produce, and corresponding experiments remain rare. Nevertheless, measurements using hyperon beams incident on target materials are crucial for advancing our understanding of nonperturbative Quantum Chromodynamics (QCD).

Theoretical models, such as the constituent quark model~\cite{cite4,cite5,cite6}, the meson-exchange picture~\cite{cite7}, and the chiral effective field theory approach~\cite{cite8,cite9,cite10,cite10a}, have studied the $\Xi N$ interaction, where $N$ represents a nucleon. The measurements of hyperon--nucleon interactions ($\Lambda p\too\Lambda p, \Sigma^{-}p\too\Sigma^{-}p/\Lambda n/\Sigma^{0} n$ and $\Sigma^{+}p\too\Sigma^{+}p$) were first carried out in the 1960s and 1970s using hyperons with momenta less than 1 GeV/$c$~\cite{cite11,cite12,cite13,cite14,cite15}. More recently, the CLAS and J-PARC E40 collaborations reported new results on $\Lambda/\Sigma$--nucleon interactions through the reactions $\Lambda p\to\Lambda p$, $\Sigma^{\pm}p\to\Sigma^{\pm}p$, and $\Sigma^{-}p\to\Lambda n$~\cite{cite16, cite17, cite18, cite19}.
In contrast, measurements related to the $\Xi N$ interaction of strangeness $S=-2$ are even more scarce.  Only a few events have been observed for each reaction~\cite{cite15,cite20,cite21,cite22,cite23,cite24,cite25}, and additional measurements are essential to better constrain theoretical models.

The $H$-dibaryon is hypothesized to be a spin and isospin singlet state of positive parity composed of six quarks ($uuddss$). It was first predicted within the bag model in the 1970s~\cite{cite26,cite27}. Since then, the $H$-dibaryon has generated considerable interest within the theoretical community~\cite{cite28,cite29,cite30}.
In particular, Refs.~\cite{cite26,cite27} predict that an $H$-dibaryon may appear as a bound state of $\Sigma\Sigma$ decaying strongly into $\Xi N$ or $\Lambda\Lambda$. Another theoretical calculation shows that the $H$-dibaryon could exist as a resonance-like state decaying into $\Lambda\Lambda$ close to the $\Lambda\Lambda$ threshold~\cite{add1}. Although the $H$-dibaryon has been searched for by many experiments, no significant signal has been found so far~\cite{cite31,cite32,cite33,cite34,cite35,cite36,cite37,cite38}. Recently, the reaction $\Xi^{0}n\too\Xi^{-}p$ has been studied~\cite{cite42}.
Similarly, the process $\Xi^{0}n\too\Lambda\Lambda$ can be explored to search for the $H$-dibaryon. The study of $\Xi N$ interactions is also helpful to understand the formation of $\Xi$ hypernuclei, on which experimental information is very scarce~\cite{cite43,cite44,cite45}.

The BESIII detector records symmetric $\EE$ collisions at the BEPCII collider~\cite{cite46}. Details of the BESIII detector can be found in Ref.~\cite{cite47}. With a sample of $(10087\pm44)\times10^{6}$ $J/\psi$ events collected by the BESIII detector~\cite{cite48}, an intense monoenergetic beam of $\Xi$ baryons can be produced through the decay $J/\psi\too\Xi^{0}\bar{\Xi}^{0}$. These $\Xi$ baryons can interact with the material of the beam pipe adjacent to the interaction point, providing a novel source for studying the $\Xi$-nucleon interaction~\cite{cite49,cite50}. The beam pipe material consists of gold ($^{196}$Au), beryllium ($^{9}$Be), and oil ($^{12}$C:$^{1}$H =1:2.13), as shown in Ref.~\cite{cite42}.

In this work, the process $\Xi^{0}n\too\Lambda\Lambda X$ is studied for the first time, where the $\Xi^{0}$ baryon is produced via $J/\psi\too\Xi^{0}\bar{\Xi}^{0}$, and the $n$ is from the $^{9}$Be, $^{12}$C, or $^{196}$Au nuclei in the beam pipe. This method has been applied in a recent study of $\Xi^{0}$-nucleus interaction at BESIII~\cite{cite42}. Since the momentum of the incident $\Xi^{0}$ is relatively high ($P_{\Xi^{0}}\approx0.818$ GeV/$c$), its interaction with atomic nuclei tends to be a direct nuclear reaction~\cite{directreaction}. In this analysis, it is assumed that the $\Xi^{0}$ interacts directly with a neutron in $^{9}$Be. To determine the cross section of $\Xi^{0}$+$^{9}$Be$\too\Lambda+\Lambda+X$ from the composite material, the reaction is further assumed to occur as a pure surface process~\cite{cite42}.

Using a {\sc geant4}-based~\cite{cite51} software package, Monte Carlo (MC) simulated event samples are produced incorporating the geometric description~\cite{cite52} of the BESIII detector and the detector response. An `inclusive' MC sample containing 10 billion generic $J/\psi$ decays is used to investigate potential backgrounds. The production of the $J/\psi$ resonance is simulated by the MC event generator {\sc kkmc}~\cite{cite53}, taking the beam-energy spread and initial-state radiation (ISR) in the $\EE$ annihilation into account. The known decay modes are generated by {\sc evtgen}~\cite{cite54,cite55} using branching fractions taken from the Particle Data Group (PDG)~\cite{cite56}, while the unknown decay modes are modeled with {\sc lundcharm}~\cite{cite57,cite58}. Final-state radiation from charged final state particles is incorporated using the {\sc photos} package~\cite{cite59}.

The signal process considered in this analysis is $J/\psi\too\Xi^0\bar{\Xi}^0$, $\Xi^0n\too\Lambda\Lambda X$ (where $X$ is typically nothing or a photon due to limited phase space for pions), $\Lambda\too p\pi^-$, $\bar{\Xi}^0\too\bar{\Lambda}\pi^0$, $\bar{\Lambda}\too \bar{p}\pi^+$ and $\pi^0\too\gamma\gamma$. To distinguish between the two $\Lambda$ candidates, the one with the higher momentum is defined as $\Lambda_{\text{H}}$, and the other as $\Lambda_{\text{L}}$. In order to determine the detection efficiency, 2.25$\times10^{6}$ signal MC events are simulated, with the angular distribution of $J/\psi\too\Xi^0\bar{\Xi}^0$ generated according to the measurement in Ref.~\cite{cite60}. The reaction $\Xi^{0}n\too\Lambda\Lambda$ is simulated under the assumption that the neutron is free, neglecting its Fermi momentum. Since the momentum of the monoenergetic incident $\Xi^{0}$ is much greater than the Fermi momentum, this approximation is reasonable. The impact of this assumption is accounted for in the systematic uncertainty evaluation. The angular distribution of the reaction process is generated using an isotropic phase-space distribution.

Charged tracks detected in the multilayer drift chamber (MDC) are required to be within a polar angle ($\theta$) range of $\left|\cos\theta\right|<0.93$, where $\theta$ is defined with respect to the $z$ axis, which is the symmetry axis of the MDC. The $\pi^{0}$ is reconstructed in its decay to $\gamma \gamma$ using electromagnetic calorimeter (EMC) showers. The deposited energy of each shower must exceed 25 MeV in the barrel region ($\left|\cos\theta\right|<0.8$) and 50 MeV in the end cap region ($0.86<\left|\cos\theta\right|<0.92$). To suppress electronic noise and showers unrelated to the event, the difference between the EMC time and the event-start time is required to be within [0, 700] ns~\cite{time1, time2}. The angle between photon candidates and all other charged tracks must be larger than $10^{\circ}$ to suppress showers originating from charged-track showers. Particle identification (PID) for charged tracks combines measurements of the specific ionization energy loss in the MDC (d$E/$d$x$) with the flight time in the time-of-flight system to form likelihoods $\mathcal{L}(h)$ ($h=p, K, \pi$) for each hadron $h$ hypothesis. Tracks are identified as protons if the proton hypothesis has the greatest likelihood
[$\mL(p)>\mL(\pi)$ and $\mL(p)>\mL(K)$], and as pions if the pion hypothesis has the greatest likelihood [$\mL(\pi)>\mL(p)$ and $\mL(\pi)>\mL(K)$].

Since the final state of the signal process is $2(p\pi^-) \bar{p}\pi^{+}2\gamma$, we require two $p$ candidates, two $\pi^-$ candidates, one $\bar{p}$ candidate, one $\pi^+$ candidate, and at least two photon candidates. For the tag side $\bar{\Xi}^0\too\bar{\Lambda}\pi^0$ with $\bar{\Lambda}\too\bar{p}\pi^{+}$, a vertex fit is performed on the $\bar{p}\pi^+$ combination.
The $\bar{\Lambda}$ signal region is defined as $|M(\bar{p}\pi^{+})-m_{\bar{\Lambda}}|<3$~MeV/$c^{2}$, where $m_{\bar{\Lambda}}$ is the nominal mass of the $\bar{\Lambda}$. Throughout this work, all nominal masses are taken from the PDG~\cite{cite56}. The invariant mass of the two photons is required to be in the $\pi^{0}$ mass window [0.11, 0.15]~GeV/$c^2$.
A 1C kinematic fit constrains the two-photon invariant mass to the nominal $\pi^{0}$ mass.
If there is more than one $\pi^{0}$ candidate in the event, only the one with the minimum value of $|M(\bar{\Lambda}\pi^0)-m_{\bar{\Xi}^0}|$ is retained. The $\bar{\Xi}^0$ signal region is defined as $M({\bar{\Lambda}\pizero})-m_{\bar{\Xi}^{0}}\in [-15, 10]$ MeV/$c^{2}$, where $m_{\bar{\Xi}^{0}}$ is the nominal $\bar{\Xi}^{0}$ mass. For the reaction $\Xi^{0}n\too\Lambda\Lambda$ with subsequent decay $\Lambda\too p\pi^{-}$,
the $\Lambda_{\text{H}}$ and $\Lambda_{\text{L}}$ candidates are formed by considering all $p\pi^-$ combinations that have disjoint $p$ and $\pi^-$ candidates in $\Lambda_{\text{H}}$ and $\Lambda_{\text{L}}$.
Vertex fits are then performed for the $(p\pi^{-})_{\text{H}}$ and $(p\pi^{-})_{\text{L}}$ combinations.
The combination with the minimum value of $(|M(p\pi^-)_{\text{H}}-m_{\Lambda}|+|M(p\pi^-)_{\text{L}}-m_{\Lambda}|)$ is retained, where $m_{\Lambda}$ is the nominal $\Lambda$ mass.

To select the signal events of $J/\psi\too\Xi^0\bar{\Xi}^0$, the invariant mass of the system recoiling against the $\bar{\Xi}^{0}$, $M_{\text{recoil}}$($\Xizerob$), is required to be in the $\Xizero$ signal region, defined as [1.295, 1.325]~GeV/$c^2$, where $M_{\text{recoil}}(\Xizerob)\equiv\sqrt{E^2_{\text{beam}}-|\vec{p}_{\bar{\Xi}^0}|^2c^{2}}/c^2$, $E_{\text{beam}}$ is the beam energy, and $\vec{p}_{\bar{\Xi}^0}$ is the measured momentum of the $\Xizerob$ candidate in the $\EE$ rest frame. The main background is $J/\psi\too\Xi^0\bar{\Xi}^0$, $\Xi^0\too\Lambda\pi^0$, $\bar{\Xi}^0\too\bar{\Lambda}\pi^0$. For this background process, $M_{\text{recoil}}(\bar{\Xi}^0\Lambda)$ is expected to be around the $\pi^0$ mass. If the missing energy $E_{\text{miss}}=E_{\text{beam}}-E_{\Lambda}$ is less than the missing momentum $|\vec{p}_{\text{miss}}|=-\sqrt{E_{\text{beam}}^2-m_{\Xi^0}^2c^4}\frac{\vec{p}_{\bar{\Xi}^0}}{|\vec{p}_{\bar{\Xi}^0}|c}-\vec{p}_{\Lambda}$, we force the recoil-mass squared to be positive, and set the recoil mass itself to be negative. $E_{\Lambda}$ and $\vec{p}_{\Lambda}$ are the energy and momentum of the $\Lambda$ in the rest frame of the $\EE$ system. To remove this background, we require $M_{\text{recoil}}(\bar{\Xi}^0\Lambda_{\text{H(L)}})<0$~GeV/$c^2$.

\begin{figure}[htbp]
\begin{center}
\includegraphics[width=0.235\textwidth]{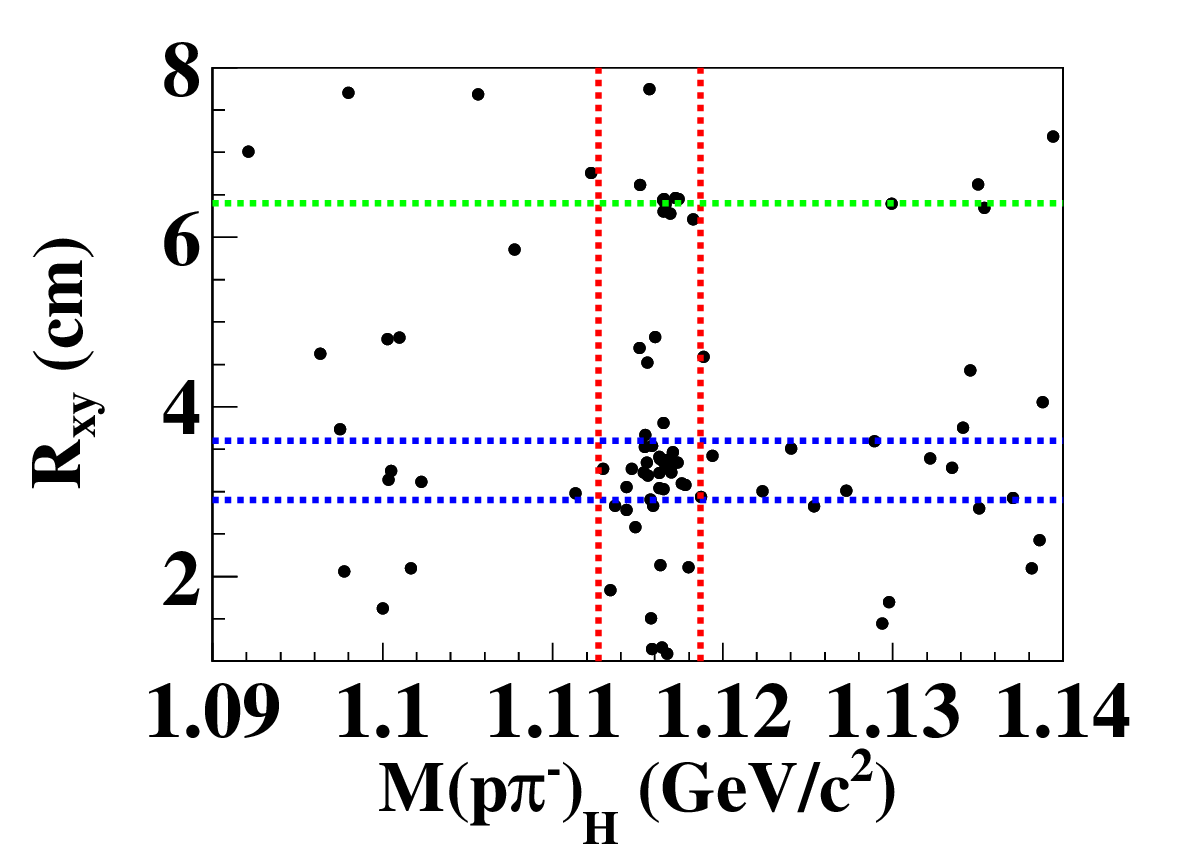}
\includegraphics[width=0.235\textwidth]{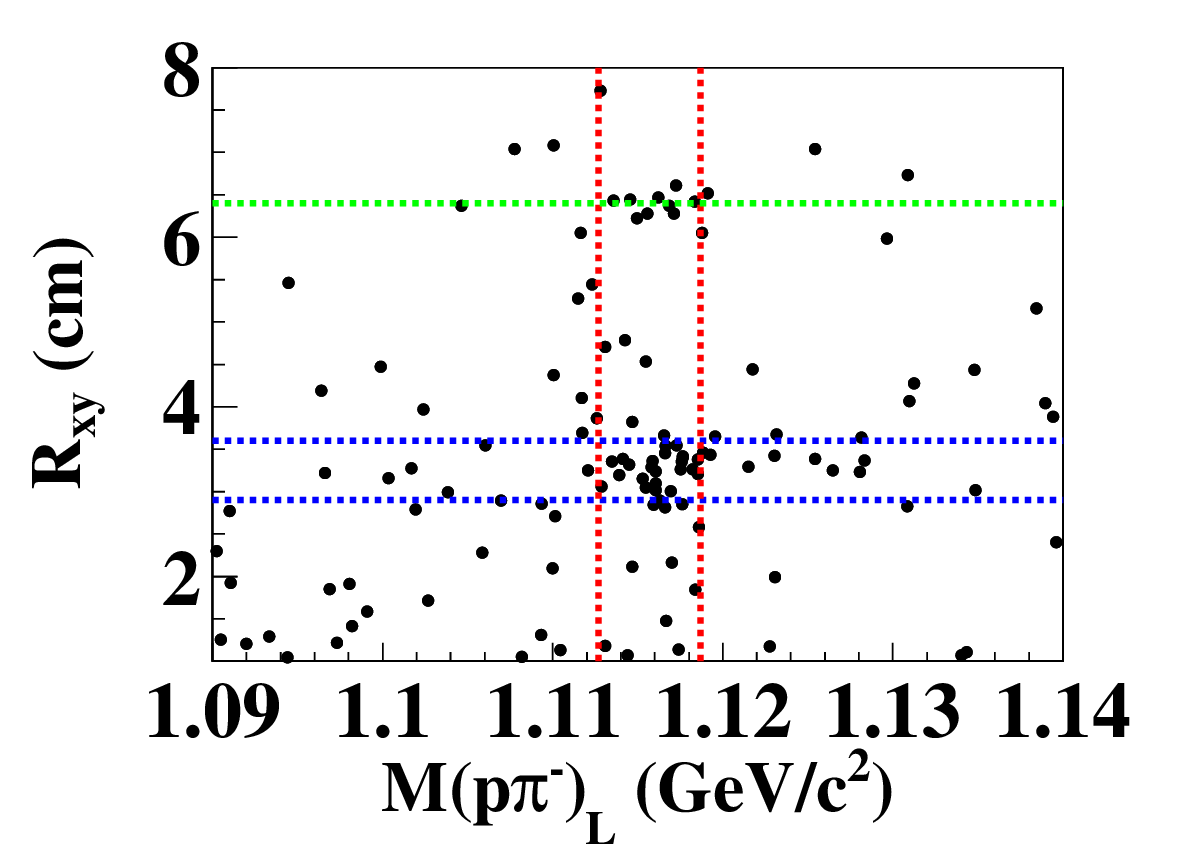}
\caption{Distribution of the transverse distance $R_{xy}$ of the $\Lambda \Lambda$ pair from the $z$ axis versus $M(p\pim)$ for data. The blue horizontal dashed lines denote the beam pipe region, the green horizontal dashed lines denote the position of the inner wall of the MDC, and the red vertical dashed lines mark the $\Lambda$ signal region.}
\label{fig:2}
\end{center}
\end{figure}

To select events where both $\Lambda$ originate from the beam pipe, we reconstruct the $\Lambda \Lambda$ vertex using a vertex fit applied to the two $\Lambda$ candidates and consider its distance $R_{xy}$ from the $z$ axis. The distribution of $R_{xy}$ versus the invariant mass $M(p\pim)$ in data is shown in Fig.~\ref{fig:2}. The beam pipe signal region is defined as $R_{xy} \in [2.9, 3.6]$~cm, taking into account the detector resolution. Clear enhancements are seen in the beam pipe and $\Lambda$ signal regions, due to $\Xi^0$ interactions with the material in the beam pipe that produce $\Lambda$ hyperons via the process $\Xizero n\too\Lambda\Lambda X$. The $\Lambda$ signal region is defined as $|M(p\pi^{-})-m_{\Lambda}|<3$~MeV/$c^{2}$, while a sideband region is defined as 9~MeV/$c^{2}<|M(p\pi^{-})-m_{\Lambda}|<15$~MeV/$c^{2}$. A cluster of events is also visible in the inner wall of the MDC region, defined as [6.0, 6.8] cm, but the signal is not statistically significant.

\begin{figure}[htbp]
\begin{center}
\includegraphics[width=0.43\textwidth]{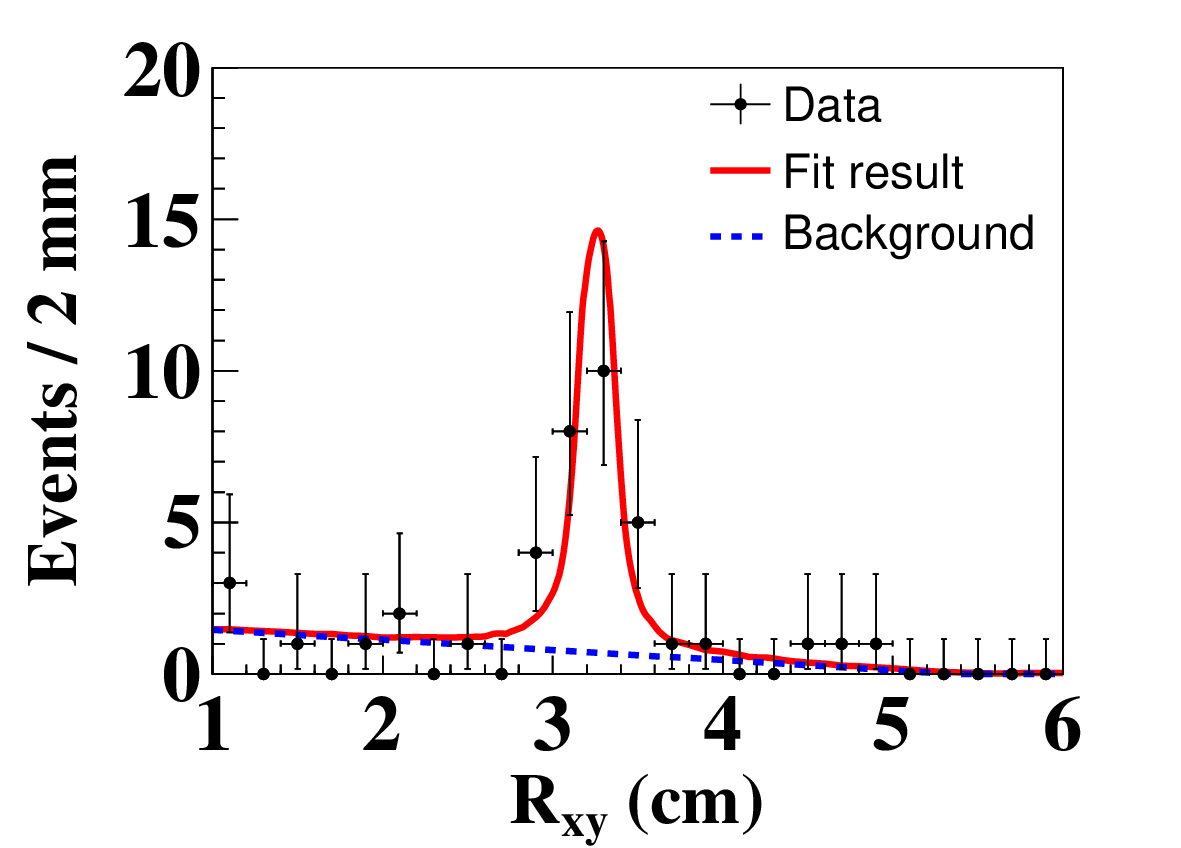}
\caption{Distribution of $R_{xy}$ in data (dots with error bars). The red solid curve is the fit result and the blue dashed curve is the background component.}
\label{fig:3}
\end{center}
\end{figure}

Figure~\ref{fig:3} shows the $R_{xy}$ distribution from data after the final event selection. It should be noted that when presenting any one $M(p\pim)$ distribution in Fig.~\ref{fig:2}, the $\Lambda$ mass window has been applied on another $M(p\pim)$ distribution. Therefore, these events in the red vertical dashed lines are same in the two $M(p\pim)$ distributions of Fig.~\ref{fig:2}, the $R_{xy}$ distribution can be obtained from any figure in Fig.~\ref{fig:2}. A clear signal corresponding to the reaction $\Xizero n\too\Lambda\Lambda X$ is observed. A detailed study of the $J/\psi$ inclusive MC sample indicates that there is no peaking background contribution in the $R_{xy}$ signal region. Additionally, no significant peak is found in the $\Lambda$ sideband events from data. To determine the signal yield, an unbinned maximum likelihood fit is performed to the $R_{xy}$ distribution. The signal shape is taken from the MC simulation, with its yield treated as a free fit parameter. The background is described by a linear function, with both the number of events and the slope allowed to vary. The fit result is shown in Fig.~\ref{fig:3}; the fitted $R_{xy}$ signal yield is $N^{\text{sig}}=24.6\pm5.9$. The statistical significance is determined to be 6.4$\sigma$ by comparing the likelihood values of the fits with and without the $R_{xy}$ signal, accounting for the change in the number of degrees of freedom. The selection efficiency for the reaction $\Xizero n\too\Lambda\Lambda X$ is $\epsilon=1.577\%$.

Using the same method as in Ref.~\cite{cite42}, the cross section of the reaction $\Xi^0+{^9\rm{Be}}\too\Lambda+\Lambda+X$ is extracted as
\begin{equation}
    \sigma(\Xi^0+{^9\rm{Be}}\too\Lambda+\Lambda+X) = \frac{N^{\rm{sig}}} {\epsilon\Br\mL_{\text{eff}}},
\end{equation}
where $\Br$ is the product of the branching ratios of all intermediate resonances, defined as $\Br\equiv\Br(\Xizerob\too\bar{\Lambda}\pizero)\Br(\bar{\Lambda}\too\bar{p}\pip)\Br(\pizero\too\gamma\gamma)\Br^{2}(\Lambda\too p\pim)$, and $\mL_{\text{eff}}$ is the effective luminosity of the $\Xizero$ flux produced from $J/\psi\too\Xizero\Xizerob$ and the
distribution of target materials, as shown in the following formula:

\begin{small}
\begin{equation}
    \mL_{\text{eff}}\!=\!\frac{\it{N}_{\it{J}/\psi}\mathcal{B}_{\it{J}/\psi}}{2+\frac{2}{3}\alpha}\!\!\!\!\int_{a}^{b}\!\!\!\!\int_{0}^{\pi}\!\!(\!1\!+\!\alpha\rm{cos}^2\theta)\!\exp(\!-\frac{\it{x}}{\rm{sin}\theta\it{\beta\gamma L}})\it{\!N\!(x)\!C\!(x)}\mathrm{d}\it{\theta}\mathrm{d}\it{x}.
\label{eq:equation2}
\end{equation}
\end{small}

In the formula for the effective luminosity, the angular distribution of the $\Xizero$ flux, the attenuation of the $\Xizero$ flux, the number of target nuclei, and the weight of different target materials are considered in turn. In the formula, $N_{J/\psi}$ is the total number of $J/\psi$ events~\cite{cite48}, $\Br_{J/\psi}$ is the branching fraction of $J/\psi\too\Xizero\Xizerob$, $\alpha$ is the parameter of the angular distribution of $J/\psi\too\Xizero\Xizerob$~\cite{cite60}, $\beta\gamma\equiv\sqrt{E^{2}_{\text{beam}}-m^{2}_{\Xizero}c^{4}}/m_{\Xizero}c^{2}$ is the ratio of the momentum and the mass of the $\Xizero$, $L\equiv c\tau$ is the decay length of the $\Xizero$, $N(x)$ is the number of target nuclei per unit volume, $a$ and $b$ are the distances between the inner and outer surface of the beam pipe and the $z$ axis, $\theta$ and $x$ are the polar angle and the distance to the $z$ axis, respectively~\cite{cite42}. The beam pipe can be regarded as infinitely long compared to the $\Xizero$ mean flight distance $\beta\gamma L$. The $C(x)$ is the cross section ratio relative to $\sigma(\Xi^0+{^9\rm{Be}}\too\Lambda+\Lambda+X)$, under the assumption that the reaction is dominated by the interaction of a $\Xizero$ baryon with a single neutron on the $^9\rm{Be}$ nucleus surface~\cite{cite61,cite62,cite63,cite64,cite65}, as discussed in Ref.~\cite{cite42}. The derivation of the formula can also be found in Ref.~\cite{cite42}. The relevant parameters are summarized in Table~\ref{tab:parameter}. The following measured cross sections are not independent as they are derived from the same basic quantities; the $\Xi$--nucleus cross sections are determined to be $\sigma(\Xi^0+{^9\rm{Be}}\too\Lambda+\Lambda+X)=(43.6\pm10.5_{\text{stat}}\pm11.1_{\text{{syst}}})$~mb, $\sigma(\Xi^0+{^{12}\rm{C}}\too\Lambda+\Lambda+X)=(48.7\pm11.7_{\text{stat}}\pm12.0_{\text{syst}})$~mb.

\begin{table}[H]
\begin{center}
\caption{Input parameters for the effective luminosity calculation using Eq. (\ref{eq:equation2}).}
\label{tab:parameter}
\begin{tabular}{ c  c}
  \hline
  \hline
  Parameter & Result   \\
  \hline
  $N_{J/\psi}$                                            & $(10087\pm44)\times10^6$  \\
  $\mathcal{B}_{J/\psi}$                                  & $(0.117\pm0.004)$\%  \\
  $\alpha$                                                & $0.514\pm0.016$  \\
  $\beta\gamma$                                           & $0.622$  \\
  $L$                                                     & $(8.69\pm0.27)$~cm  \\
  $N(x)$ & $\begin{cases}
    5.91\times10^{22}~\rm{cm^{-3}}, 3.148564~\rm{cm}\leq \it{x}\leq \rm{3.15}~\rm{cm} \\
    1.24\times10^{23}~\rm{cm^{-3}}, 3.15~\rm{cm}< \it{x}\leq \rm{3.23}~\rm{cm} \\
    3.45\times10^{22}~\rm{cm^{-3}}, 3.23~\rm{cm}< \it{x}\leq \rm{3.31}~\rm{cm}\\
    1.24\times10^{23}~\rm{cm^{-3}}, 3.31~\rm{cm}< \it{x}\leq \rm{3.37}~\rm{cm} \\
  \end{cases}$
  \\
  $C(x)$     & $\begin{cases}
    8.437, \ \ \ 3.148564~\rm{cm}\leq \it{x}\leq \rm{3.15}~\rm{cm} \\
    1, \ \ \ \ \ \ \ \   3.15~\rm{cm}< \it{x}\leq \rm{3.23}~\rm{cm} \\
    1.090, \ \ \  3.23~\rm{cm}< \it{x}\leq \rm{3.31}~\rm{cm} \\
    1, \ \ \ \ \ \ \ \ \  3.31~\rm{cm}< \it{x}\leq \rm{3.37}~\rm{cm} \\
  \end{cases}$ \\

  \hline
  \hline
\end{tabular}
\end{center}
\end{table}

\begin{figure}[htbp]
\begin{center}
\includegraphics[width=0.5\textwidth]{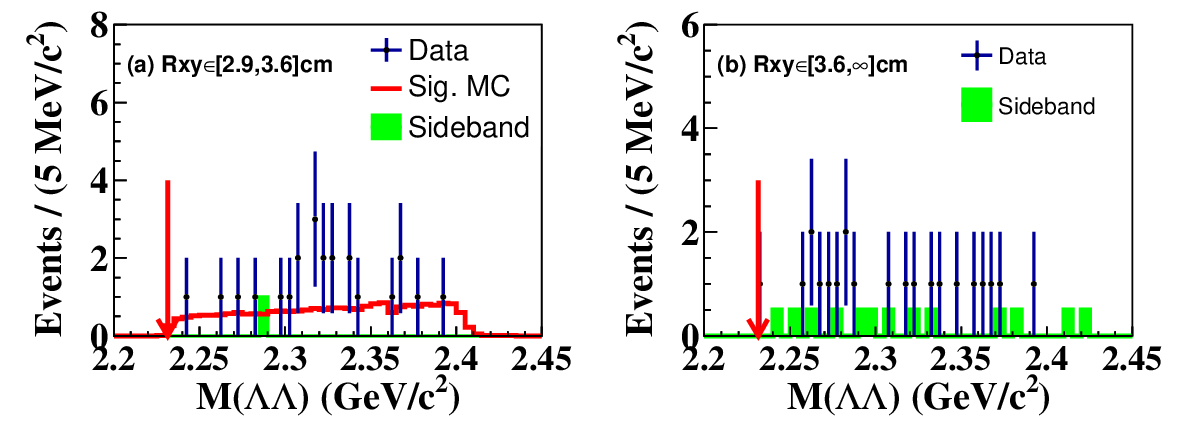}
\caption{Distributions of $M(\Lambda\Lambda$) for data in the regions $2.9<R_{xy}<3.6$~cm (a) and $R_{xy}>3.6$~cm (b). The green shaded histograms correspond to the normalized events from the 2D sideband region of $\Lambda_{\text{H}}$ and $\Lambda_{\text{L}}$, while the red line corresponds to the signal MC distribution of the scattering process that is normalized by the total number of events for data. The red arrows mark the $\Lambda\Lambda$ mass threshold. }
\label{fig:4}
\end{center}
\end{figure}

We also search for an $H$-dibaryon signal in the $\Lambda\Lambda$ final state. Since the mean lifetime of the $H$-dibaryon is unknown, it could decay either within the beam pipe region or after traveling some distance.
Figure~\ref{fig:4} shows the $M(\Lambda\Lambda$) distributions for events with two selected $\Lambda$ candidates, both inside and outside the beam pipe region. It is worth mentioning that the efficiency near the $\Lambda\Lambda$ threshold is about half of the average efficiency, so even if the $H$-dibaryon appears near the $\Lambda\Lambda$ threshold, our method can still detect it. Based on the available statistics, no obvious peaks are observed in either $M(\Lambda\Lambda$) distributions. Therefore, no significant short-lifetime or long-lifetime $H$-dibaryon signal is found in the process $\Xizero n\too\Lambda\Lambda X$.

The sources of systematic uncertainties related to the measurement of cross sections are discussed in the following.

The uncertainties due to tracking, PID and photon reconstruction are each 1\% per track or photon, respectively~\cite{cite66, cite66b}. The uncertainty from the requirement on the number of tracks is studied using the control sample of $J/\psi\too\Xim\Xip\too\LLB\pip\pim\too p\bar{p}\pip\pim\pip\pim$ decays. The uncertainties from $\Lambda/\bar{\Lambda}/\Xi^0/\bar{\Xi}^0$ mass windows are evaluated with the Barlow test method~\cite{cite67}, where the largest difference relative to the nominal result is taken as the uncertainty. Similarly, the uncertainty from the $M_{\text{recoil}}(\bar{\Xi}^0\Lambda)$ requirement is also evaluated with the Barlow test method, and found to be negligible.

In the MC simulation of the signal process, the neutron in the nucleus is assumed to be at rest. However, due to the presence of Fermi momentum, a difference arises between the momentum distribution of the $\Lambda\Lambda$ system, $P(\Lambda\Lambda)$, in data and MC. Since the monoenergetic momentum of the incident $\Xizero$ is much larger than the Fermi momentum, this effect is expected to be small. To estimate the uncertainty from $P(\Lambda\Lambda)$, the momentum of the free neutron is varied by $\pm 0.1$ GeV/$c$ along the incident $\Xizero$ direction in the generated signal MC, and the largest deviation from the nominal result is taken as the uncertainty.

The distribution of the $\Lambda\Lambda$ invariant mass, $M(\Lambda\Lambda$), is almost flat in data from 2.23 to 2.40 GeV/$c^{2}$, consistent with expectations when accounting for the influence of Fermi momentum. The center-of-mass energy of the reaction system is adjusted to match the $M(\Lambda\Lambda)$ distribution between data and MC simulation, as shown in Fig.~\ref{fig:4} (left). The systematic uncertainty from the $M(\Lambda\Lambda)$ distribution is obtained by comparing the efficiencies of the nominal signal MC and the weighted signal MC, where the latter is reweighted to match the signal distribution in data. The reaction $\Xizero n\too\Lambda\Lambda$ is simulated with a uniform angular distribution over the phase space to estimate the nominal efficiency. The correction of signal events is derived from data, and the difference between the nominal and weighted efficiencies is taken as the uncertainty.

The uncertainty from the contribution of the reaction $\Xi^{0}n\too\Lambda\Sigma^0$ is estimated by generating a signal MC sample above the mass threshold of $\Lambda\Sigma^0$ and determining the weighted efficiency under isospin conservation, where $\sigma(\Xi^{0}n\too\Lambda\Lambda)=\sigma(\Xi^{0}n\too\Lambda\Sigma^0)$. The difference between the nominal and weighted efficiencies is taken as the uncertainty.

The uncertainty from the MC statistics is estimated according to the number of generated signal MC events. The uncertainty from the fit procedure includes contributions from the signal shape, the fit range and the background shape. The uncertainty from the signal shape is evaluated by comparing the nominal fit with an alternative fit that uses the MC-determined signal shape convolved with a free Gaussian function. The uncertainty from the fit range is obtained by varying the limit of the fit range by $\pm0.5$ cm, and the uncertainty associated with the background shape is estimated by replacing the first-order polynomial function with a constant. The uncertainty on the total number of $J/\psi$ events is estimated in Ref.~\cite{cite48}, and the uncertainty of the branching fractions is taken from the PDG~\cite{cite56}. The uncertainty from the position of the $\EE$ interaction point is obtained by changing the integral range by $\pm$0.1 cm, and the largest difference in the result is taken as the corresponding uncertainty. To estimate the uncertainty from the assumption of the cross section ratio, we adopt an extreme scenario in which the cross section is proportional to the number of neutrons in the nucleus, and the difference between the nominal and the extreme assumption is taken as the uncertainty. Finally, we vary the angular distribution parameter $\alpha_{J/\psi\too\Xizero\Xizerob}$ and the $\Xizero$ mean lifetime by $\pm1\sigma$ to estimate systematic uncertainties; this is found to be negligible.

All the systematic uncertainties are assumed to be independent and are combined in quadrature to obtain the total systematic uncertainty, as shown in Table~\ref{tab:sumerror}.
\begin{table}[htbp]
\begin{center}
\caption{Summary of the relative systematic uncertainty (in $\%$); $\sigma_{\text{Be}}$ represents the cross section $\sigma(\Xi^0+{^9\rm{Be}}\too\Lambda+\Lambda+X)$, and $\sigma_{\text{C}}$ represents the cross section $\sigma(\Xi^0+{^{12}\rm{C}}\too\Lambda+\Lambda+X)$.}
\label{tab:sumerror}
\begin{tabular}{ c c c}
  \hline
  \hline
  Source & $\sigma_{\text{Be}}$ & $\sigma_{\text{C}}$  \\
  \hline
  Track efficiency                                           & 6.0 & 6.0 \\
  Photon efficiency                                          & 2.0 & 2.0 \\
  PID efficiency                                             & 6.0 & 6.0 \\
  Track number                                               & 3.0 & 3.0 \\
  Mass window                                                & 6.8 & 6.8 \\
  $(\Lambda+\Lambda)$ momentum                               & 9.0 & 9.0 \\
  $M(\Lambda\Lambda)$ distribution                           & 1.3 & 1.3 \\
 \ \ Angular distribution of $\Xi^{0}n\too\Lambda\Lambda$ \ \       & \ \ 16.3 \ \ & \ \ 16.3 \ \ \\
  $\Xi^0n\too\Lambda\Sigma^0$                                & 9.3 & 9.3 \\
  MC statistics                                              & 0.1 & 0.1 \\
  Fit procedure                                              & 1.8 & 1.8 \\
  Total number of $J/\psi$ events      & 0.4 & 0.4 \\
  Branching fractions                                        & 3.7 & 3.7 \\
  $e^+e^-$ interaction point                                 & 2.7 & 2.7 \\
  $\Xi^0$ mean life time                                     & 2.8 & 2.8 \\
  Ratio of cross section for nuclei                          & 7.1 & 2.3 \\
  \hline
  Sum                                                        & 25.5 & 24.6 \\
  \hline
  \hline
\end{tabular}
\end{center}
\end{table}

In summary, the inelastic scattering $\Xi^0+{^9\rm{Be}}\too\Lambda+\Lambda+X$ is observed with a statistical significance of 6.4$\sigma$ at BESIII, where $\Xizero$ is from the process $J/\psi\too\Xizero\Xizerob$ and $n$ is from the material in the beam pipe. The cross section is measured to be $\sigma(\Xi^0+{^9\rm{Be}}\too\Lambda+\Lambda+X)=(43.6\pm10.5_{\text{stat}}\pm11.1_{\text{syst}})$ mb at $P_{\Xi^0}\approx0.818$ GeV/$c$. If the effective number of reactive neutrons in a $^9\rm{Be}$ nucleus is taken as 3~\cite{cite18}, the cross section of $\Xizero n\too\Lambda\Lambda X$ for a single neutron is determined to be $\sigma(\Xi^{0}n\too\Lambda\Lambda X)=(14.5\pm3.5_{\text{stat}}\pm3.7_{\text{syst}})$~mb, which is consistent with theoretical predictions in Refs.~\cite{cite21,cite24,cite25}. The measured reaction $\Xizero n\too\Lambda\Lambda X$ is mainly composed of the reactions $\Xizero n\too\Lambda\Lambda$ and $\Xizero n\too\Lambda\Sigma^0\too\Lambda\Lambda\gamma$. Furthermore, we do not observe any significant $H$-dibaryon signal in the $\Lambda\Lambda$ final state for this reaction process.

\section{ACKNOWLEDGMENTS}
The BESIII Collaboration thanks the staff of BEPCII (https://cstr.cn/31109.02.BEPC) and the IHEP computing center for their strong support. This work is supported in part by National Key R\&D Program of China under Contracts Nos. 2023YFA1606000, 2023YFA1606704; National Natural Science Foundation of China (NSFC) under Contracts Nos. 12375071, 11635010, 11935015, 11935016, 11935018, 12025502, 12035009, 12035013, 12061131003, 12192260, 12192261, 12192262, 12192263, 12192264, 12192265, 12221005, 12225509, 12235017, 12361141819; Natural Science Foundation of Henan under Contract No. 242300421163; the Chinese Academy of Sciences (CAS) Large-Scale Scientific Facility Program; CAS under Contract No. YSBR-101; 100 Talents Program of CAS; The Institute of Nuclear and Particle Physics (INPAC) and Shanghai Key Laboratory for Particle Physics and Cosmology; ERC under Contract No. 758462; German Research Foundation DFG under Contract No. FOR5327; Istituto Nazionale di Fisica Nucleare, Italy; Knut and Alice Wallenberg Foundation under Contracts Nos. 2021.0174, 2021.0299; Ministry of Development of Turkey under Contract No. DPT2006K-120470; National Research Foundation of Korea under Contract No. NRF-2022R1A2C1092335; National Science and Technology fund of Mongolia; Polish National Science Centre under Contract No. 2024/53/B/ST2/00975; STFC (United Kingdom); Swedish Research Council under Contract No. 2019.04595; U. S. Department of Energy under Contract No. DE-FG02-05ER41374.

\end{document}